\documentclass{PoS}

\usepackage{amsmath}
\usepackage{amsfonts}
\usepackage{amssymb}
\usepackage{latexsym}
\usepackage{epsf}

\newcommand{\be}{\begin{equation}}
\newcommand{\ee}{\end{equation}}
\newcommand{\bi}{\begin{itemize}}
\newcommand{\ei}{\end{itemize}}
\newcommand{\bea}{\begin{eqnarray}}
\newcommand{\eea}{\end{eqnarray}}

\let\a=\alpha \let\b=\beta    
        \let\l=\lambda
\let\m=\mu    \let\n=\nu

\let\==\equiv

\newcommand{\cO}{\mathcal{O}}
\newcommand{\cP}{\mathcal{P}}
\newcommand{\cR}{\mathcal{R}}

\newcommand{\unit}{{\bf{1}}}

\newcommand{\half}{\tfrac{1}{2}}

\newcommand{\cb}{\bar{c}}
\newcommand{\gb}{\bar{g}}

\newcommand{\p}{\partial}

\newcommand{\Db}{\bar{D}}

\newcommand{\Rb}{\bar{R}}

\newcommand{\Gt}{\tilde{G}}
\newcommand{\Lt}{\tilde{\Lambda}}

\title{Higher Derivative Gravity \\ from the Universal Renormalization Group Machine}

\ShortTitle{Higher Derivative Gravity from the URGM}

\author{Kai Groh, Stefan Rechenberger, \speaker{Frank Saueressig} and Omar Zanusso\\
        Institute of Physics, University of Mainz\\
        Staudingerweg 7, D-55099 Mainz, Germany \\
        E-mail: \email{kgroh@thep.physik.uni-mainz.de}, 
        \email{rechenbe@thep.physik.uni-mainz.de},
\email{saueressig@thep.physik.uni-mainz.de}, 
\email{zanusso@thep.physik.uni-mainz.de}}

\abstract{
We study the renormalization group flow of higher derivative gravity, utilizing the functional renormalization group equation for the average action. Employing a recently proposed algorithm, termed the universal renormalization group machine, for solving the flow equation, all the universal features of the one-loop beta-functions are recovered. While the universal part of the beta-functions admits two fixed points, we explicitly show that the existence of one of them depends on the choice of regularization scheme, indicating that it is most probably unphysical.}

\FullConference{The 2011 Europhysics Conference on High Energy Physics-HEP 2011,\\
		July 21-27, 2011\\
		Grenoble, Rh\^{o}ne-Alpes France}

\begin{document}
\section{Introduction}
Finding a consistent and predictive UV completion of gravity is one of the most challenging tasks in theoretical high energy physics to date. 
In this endeavor it soon became apparent that understanding the theory's renormalization group (RG) flow may be a crucial ingredient.
The emphasis is thereby on fixed points (FPs), which could provide a consistent UV completion of the theory within 
Wilson's formulation of renormalization. In this context, it is natural to distinguish the following cases: firstly, there may be a Gaussian fixed point (GFP) linked to the free theory. This is the structure underlying the asymptotic freedom of a perturbatively renormalizable theory. Secondly, the flow may possess non-Gaussian fixed points (NGFPs) where the corresponding fundamental action contains interactions.  A theory, whose UV completion is provided by such a NGFP is termed asymptotically safe. Notably, asymptotic safety may be as predictive as asymptotic freedom \cite{Percacci:2011fr}.

Already at a very early stage, it was observed that the UV-completion of the Einstein-Hilbert action is not given by a GFP, i.e., the perturbative 
quantization procedure does not lead to an asymptotically free quantum field theory  \cite{Goroff:1985th}.
In order to rescue perturbative renormalizability the Einstein-Hilbert action was complemented by introducing fourth order operators \cite{Stelle:1976gc}.
This improves the UV behavior of the theory and the marginal couplings associated with the four-derivative terms are asymptotically free at the one loop level
\cite{Julve:1978xn, deBerredoPeixoto:2004if}. Unfortunately, this improvement comes at the price of introducing massive negative norm states \cite{Stelle:1977ry, Barth:1983hb}, so that it is commonly believed that higher derivative gravity is not unitary.

Along a different line Weinberg proposed that gravity could be asymptotically safe \cite{Weinberg:1980gg}. A key ingredient in investigating this possibility is the gravitational version of the Wetterich equation \cite{Reuter:1996cp}, which allows to investigate non-perturbative properties of the gravitational RG flow.
Since its advent this tool provided an impressive body of evidence that gravity indeed possesses a suitable NGFP \cite{1999-2007}. Recently, these developments have culminated in the proposal of  a systematic algorithm for solving the gravitational functional renormalization group equation (FRGE) using off-diagonal heat-kernel techniques: the universal RG machine (URGM) \cite{Benedetti:2010nr}.

In the sequel we will rederive the perturbative $\beta$-functions of higher derivative gravity \cite{Julve:1978xn, deBerredoPeixoto:2004if} from the FRGE \cite{Reuter:1996cp}. Similar studies have been carried out before \cite{Codello:2006in,Niedermaier:2009zz}, where it was observed that keeping track of the quadratic and quartic divergences has a drastic effect on the fixed point structure of the RG flow. These contributions shift the fixed point for Newton's constant and cosmological constant to non-zero values, rendering the theory asymptotically safe instead of asymptotically free. The main purpose of the present work is the demonstration that the URGM
 recovers these results. Surprisingly, the regularization scheme intrinsic to the URGM unveils certain features in the fixed point structure of the theory, that have not been stressed before. 

\section{The flow equation for higher-derivative gravity}
In the case of higher-derivative gravity, our ansatz for the average action $\Gamma_k$ 
contains all gravitational interaction monomials with four or less powers of momentum:\footnote{We shall neglect total derivative terms.}
\be\label{ansatz}
\Gamma_k = \int d^4x \sqrt{g} \left[ 2 Z_k \Lambda_k - Z_k R + \frac{1}{2\lambda_k} C^2 - \frac{\omega_k}{3 \lambda_k} R^2 + \frac{\theta_k}{\lambda_k} E \right] + S_{\rm GF} + S_{\rm c} + S_{\rm b} \, .
\ee
Here $C^2 \equiv C_{\m\n\a\b} C^{\m\n\a\b}$ abbreviates the square of the Weyl tensor,
$E = C^2 - 2 R_{\m\n} R^{\m\n} + \tfrac{2}{3} R^2$ is the integrand of the Euler topological invariant,
$Z \equiv 1/16 \pi G$ contains the dimensionful Newton's constant $G$, and all coupling constants are allowed to depend on
the RG scale $k$.

In order to consistently quantize the theory, we employ the background field method to fix the diffeomorphism invariance,
 splitting the averaged metric $g_{\m\n} = \gb_{\m\n} + h_{\m\n}$ into a fixed (but arbitrary) background metric $\gb_{\m\n}$ and 
fluctuations $h_{\m\n}$. The gauge-fixing action $S_{\rm GF}$ is then taken of the form
\be
S_{\rm GF} = \half \int d^4x \sqrt{\gb} \, F_\mu \, Y^{\m\n} \, F_\n \, ,
\ee
where $F_\m = \Db^\n h_{\m\n} - \eta \Db_\n h$ and the bar denotes covariant derivatives with respect to the background metric. Since 
the gravitational part of the action contains terms with up to four derivatives of the fluctuation fields, we also allow for four derivatives in $S_{\rm GF}$, employing 
the minimal gauge \cite{deBerredoPeixoto:2004if}
\be\label{Ymin}
Y^{\m\n} = \lambda^{-1} \left[ \gb^{\m\n} \, \Delta +  \sigma^{\rm b} \, \Db^\m \Db^\n + V^{\m\n}_{\rm b} \right] \, ,
\ee
where $\Delta \equiv - \Db^2$ and the gauge parameters are given by
%
$\eta = \tfrac{1}{4} \tfrac{1+4\omega}{1+\omega}$, $\sigma^{\rm b} = \tfrac{1-2\omega}{3}$, and $V^{\m\n}_{\rm b} = \Rb^{\m\n}$, respectively.
%
This choice has the virtue of removing all non-minimal four-derivative terms (as, e.g., $\Delta \Db^\m \Db^\n$) from the Hessian of the gravitational fluctuations.

This type of higher-derivative gauge-fixing results in two ghost-terms which take into account the Faddeev-Popov determinant \cite{Barth:1983hb}. 
The operator $F_\mu$ leads to a complex pair of ghosts $\cb, c$ with action
\be\label{c-ghost}
S_{\rm c} = \int d^4x \sqrt{\gb} \, \bar{c}_\m \, \left[ \Delta \delta^\m_\n
-\tfrac{1-2\omega}{2(1+\omega)} \, 
\Db^\m \Db_\n
-\Rb^{\m}{}_{\n}
\right] \, c^\n \, ,
\ee
while the contribution of $Y^{\m\n}$ is captured by a third (real) ghost field $b$
\be\label{b-ghost}
S_{\rm b} = \half \int d^4x \sqrt{\gb} \, b_\m \, Y^{\m\n} \, b_\n \, .  
\ee

The key ingredient for deriving the $\beta$-functions 
controlling the scale-dependence of the coupling constants
contained in the ansatz \eqref{ansatz} is the FRGE for the 
gravitational average action \cite{Reuter:1996cp}
\be\label{FRGE}
\p_t \Gamma_k[\Phi, \bar{\Phi}] = \half {\rm STr} \left[ \left( \frac{\delta^{2} \Gamma_k}{\delta \Phi^A \delta \Phi^B } + \cR_k \right)^{-1} \, \p_t \cR_k  \right] \, .
\ee
Here, $t = \log(k/k_0)$, STr contains a minus sign for Grassmann-valued fields, and $\Phi = \{ h_{\m\n},\bar{c},c,b\}$ and $\bar{\Phi}$ denote the collection of fluctuation and background fields, respectively.
Moreover, $\cR_k(p^2)$ is a (matrix-valued) infrared cutoff
which provides a $k$-dependent mass term for fluctuations with momenta $p^2<k^2$. The interplay between the regulated propagator and the derivative of the regulator 
thereby ensures that the trace remains finite for all values of $k$. In constructing the cutoff we follow the URGM and  choose $\cR_k$ in such a way that it provides a mass term to the highest power of the Laplacians appearing in the kinetic terms $\Delta \mapsto P_k(\Delta) \equiv \Delta + R_k(\Delta/k^2)$.\footnote{In the terminology of \cite{cpr}, this constitutes a cutoff of Type I.} The profile function is taken as $R_k(p^2) = (k^2 - p^2) \theta(k^2-p^2)$. 
 
Upon substituting the ansatz \eqref{ansatz}, the trace in \eqref{FRGE} splits into a gravitational and two ghost parts, 
$\p_t \Gamma_k = T^{\rm grav} + T^{\rm c} + T^{\rm b}$, where
\be\label{ERGE}
\begin{split}
T^{\rm grav} \equiv & 
\half {\rm Tr} \left[ \left( \frac{\delta^2(\Gamma_k^{\rm grav} + S_{\rm GF})}{\delta h \delta h} + \cR_k^{\rm grav} \right)^{-1} \p_t \cR_k^{\rm grav} \right]\, , \\
 T^{\rm b} \equiv & \,  - \half {\rm Tr} \left[ \left( \frac{\delta^2 S_{\rm b}}{\delta b \delta b} + \cR_k^{\rm b} \right)^{-1} \p_t \cR_k^{\rm b} \right]\, , \qquad 
T^{\rm c} \equiv - {\rm Tr} \left[ \left( \frac{\delta^2 S_{\rm c}}{\delta \bar{c} \delta c } + \cR_k^{\rm c} \right)^{-1} \p_t \cR_k^{\rm c} \right] \, .
\end{split}
\ee
We shall now evaluate these traces employing the off-diagonal heat-kernel methods advocated in \cite{Anselmi:2007eq,Benedetti:2010nr,Codello:2011yf}. In this course, we neglect the $k$-dependence of all coupling constants inside the traces, which corresponds to the one-loop approximation of the flow equation.

We start with the gravitational trace. Abbreviating pairs of symmetric (external) tensor indices with a single label $i,j$, i.e., $h_i \equiv h_{\m\n}$, etc.\ and using the variations \cite{Barth:1983hb}, the part of the action quadratic in the metric fluctuations takes the form\footnote{At this stage it is consistent to drop all terms containing derivatives of curvatures, since these do not carry any information about the flow of the coupling constants contained in the ansatz \eqref{ansatz}.} \cite{deBerredoPeixoto:2004if}
$\delta^2 \left[ \half \Gamma_k^{{\rm grav}}+S_{\rm GF} \right] = \half h^{i} \left[ K_{ij} \Delta^2 + D^{(\rho\sigma)}_{ij} \Db_\rho \Db_\sigma + W_{ij} \right] h^{j}$. 
 Implementing the prescription of the Type I cutoff detailed above fixes
$\cR_k^{\rm grav} = K_{ij} (P_k(\Delta)^2 - \Delta^2)$. In order to proceed further, we note that the matrix $K$ is easily inverted. Defining $[V^{(\rho\sigma)}]^{i}{}_{j} \equiv [K^{-1}]^{il} D^{(\rho\sigma)}_{lj}$ and $[U]^{i}{}_{j} \equiv [K^{-1}]^{il} W_{lj}$, the quadratic fluctuations become
\be\label{gravquad2}
\delta^2 \left[ \half \Gamma_k^{{\rm grav}}+S_{\rm GF} \right] = \half h^{i} K_{i}{}^{l} \left[ \unit_{lj}\Delta^2 + V^{(\rho\sigma)}_{lj} \Db_\rho \Db_\sigma + U_{lj} \right] h^{j} \, . 
\ee
Since the ``interaction vertices'' $V$ and $U$ are of mass-dimension two and four, respectively, the inverse of the regulated propagator can be constructed perturbatively in $U$ and $V$. Neglegting all interactions of mass-dimension six or higher, we obtain
\be\label{Tgrav}
T^{\rm grav} \simeq 
{\rm Tr} \left[\tfrac{\p_t P_k}{P_k}\right]
 -
 {\rm Tr} \left[U \tfrac{\p_t P_k}{P^3_k}\right]
 -
 {\rm Tr} \left[V^{(\mu\nu)} \Db_\mu \Db_\nu \tfrac{\p_t P_k}{P^3_k}\right]
 +
 {\rm Tr} \left[V^{(\mu\nu)}V^{(\alpha\beta)}\Db_\mu \Db_\nu \Db_\alpha \Db_\beta \tfrac{\p_t P_k}{P^5_k}\right] \, ,
\ee 
where the trace also contains a summation over internal indices. Evaluating the traces utilizing the off-diagonal heat-kernel \cite{Anselmi:2007eq,Benedetti:2010nr} yields
\be
\begin{split}
T^{\rm grav} = 
\frac{1}{(4\pi)^2}\int d^4x \sqrt{g} & \, 
 \Bigl[ 10 k^4 + k^2 \left(\tfrac{10}{3} R + \tfrac{1}{6} V_i{}^i{}_\mu{}^\mu\right) 
 +\tfrac{5}{18}R^2-\tfrac{1}{9}R_{\mu\nu}R^{\mu\nu}-\tfrac{8}{9}R_{\mu\nu\alpha\beta}R^{\mu\nu\alpha\beta}\\
 & -\tfrac{1}{6} R_{\mu\nu}V_i{}^i{}^{\mu\nu} 
 +\tfrac{1}{12}RV_i{}^i{}_\mu{}^\mu - U_i{}^i
 +\tfrac{1}{48}V_{ij}{}_\mu{}^\mu V^{ji}{}_\nu{}^\nu +\tfrac{1}{24}V_{ij}{}_{\mu\nu} V^{ji}{}^{\mu\nu}
 \Bigr]\, ,
\end{split}
\ee
which encompasses the well-known result for the (cutoff-independent) four-derivative terms. The gravitational trace is found by substituting the properly adjusted and rather lengthy matrices $U$ and $V$ of \cite{deBerredoPeixoto:2004if}.

The inverse (regularized) propagators appearing in the ghost traces $T^{\rm c}$ and $T^{\rm b}$ have the form
\be\label{g2reg}
 [\Gamma^{(2)}_k + \cR_k]_\m{}^\n = \delta_\m^\n P_k(\Delta) + \sigma \Db_\m \Db^\n + V_\m{}^\n \equiv \cP + V_\m{}^\n \, ,
\ee
where $\sigma$ is a fixed parameter depending on the coupling $\omega$ and $V_\m{}^\n$ is an endomorphism proportional to the Ricci tensor.
The inverse of such an operator can be constructed explicitly as a power series of the curvature and endomorphism tensors by
generalizing the techniques \cite{Barvinsky:1985an}. At zeroth order in the curvature, an explicit computation establishes
\be\label{prop}
[\cP^{-1}_0]_\m{}^\n =  \frac{1}{P_k} \delta_\m^\n - \sigma \Db_\m \Db^\n \frac{1}{P_k (P_k - \sigma \Delta)} \, .
\ee 
The curvature corrections to this expression can be found order by order in a systematic bootstrap calculation. Including all terms up to $\cO(R^2)$ the result is
\be
\begin{split}
[\cP^{-1}_2&\,]_\m{}^\n =   [\cP^{-1}_0]_\m{}^\a \Big[ \delta_\a{}^\n  + \sigma \, \big(  [P_k , \Db_\a \Db^\n ] + \sigma \Db_\a [\Db^\n, \Delta] \big) \, \frac{1}{P_k(P_k - \sigma \Delta)} \\
& + \sigma^2 \, \big(  [P_k , \Db_\a \Db^\b ] + \sigma \Db_\a [\Db^\b, \Delta] \big) \, 
\big(  [P_k , \Db_\b \Db^\n ] + \sigma \Db_\b [\Db^\n, \Delta] \big) \, \frac{1}{P_k^2(P_k - \sigma \Delta)^2} \Big] \, . \\
\end{split}
\ee

With this result at hand, it is now straightforward to evaluate a generic ghost trace of the form $T^{\rm gh} \equiv {\rm Tr} [\cP + V]^{-1} \p_t \cR_k^{\rm gh}$. Expanding in $V$ and substituting explicit expressions for the commutators, the URGM yields
\be\label{genvectrace}
\begin{split}
T^{\rm gh} = & \, \frac{1}{(4\pi)^2} \int d^4x \sqrt{g} \Big[  \, 
k^4 \left(3 - \tfrac{2}{\sigma} - \tfrac{2}{\sigma^2} \log(1-\sigma) \right) 
+ k^2 R \left( 1 + \half \psi - \tfrac{1}{3\sigma} \log(1-\sigma) \right) \\ 
& \, - k^2 V \left(\tfrac{3}{4} + \tfrac{1}{2\sigma(1-\sigma)} + \tfrac{1}{2\sigma^2} \log(1-\sigma) \right) 
 - \tfrac{11}{90} R_{\m\n\a\b} R^{\m\n\a\b} \\
& \, + \left( \tfrac{1}{12} \psi^2 + \tfrac{1}{6} \psi - \tfrac{2}{45} \right) R_{\m\n} R^{\m\n} 
+ \left( \tfrac{1}{24} \psi^2 + \tfrac{1}{6} \psi + \tfrac{1}{9} \right) R^2 
 - \left( \tfrac{1}{6} \psi^2 + \tfrac{2}{3} \psi \right) R_{\m\n} V^{\m\n}  
\\
&
 - \left(\tfrac{1}{12} \psi^2 + \tfrac{1}{6} \psi + \tfrac{1}{3} \right) R V
 + \left( 1 + \half \psi + \tfrac{1}{12} \psi^2 \right) V_{\m\n} V^{\m\n}  + \tfrac{1}{24} \psi^2 V^2
\Big] \, ,
\end{split}
\ee
where $\psi \equiv \sigma/(1-\sigma)$ and $V \equiv V_\m{}^\m$. Notably, the universal four-derivative piece agrees with earlier results on the $B_4$ heat-kernel coefficient 
of the non-minimal differential operator \eqref{g2reg} (with $P_k(\Delta) = \Delta$), providing an independent verification of earlier results \cite{Barvinsky:1985an}. Moreover, one can check explicitly, that the  $\sigma$-dependence is such that the limit $\sigma \rightarrow 0$ is smooth
and the resulting trace reduces to the one for the corresponding minimal differential operator. The explicit expressions for $T^{\rm c}$ and $T^{\rm b}$ are then obtained from \eqref{genvectrace} by taking into account the appropriate prefactors and substituting
\be\label{parameters}
\sigma^{\rm b} = \tfrac{1}{3} (1-2\omega) \, , \quad V_{\rm b}^{\m\n} = \Rb^{\m\n} \, , \qquad \sigma^{\rm c} = -\tfrac{1-2\omega}{2(1+\omega)} \, , \quad  \, V_{\rm c}^{\m\n} = - \Rb^{\m\n} . 
\ee

Combining the three traces \eqref{ERGE}, we obtain the final form of the flow equation
\be\label{floweq}
\begin{split}
\p_t \Gamma_k =  \frac{1}{(4\pi)^2} & \int d^4x\sqrt{g} \bigg[  \frac{133}{20} C^2 - \frac{196}{45} E +  \frac{5}{36} \left(1 + 8 \omega + 12 \omega^2 \right) R^2  \\ &
- \left\{ \tfrac{Z \lambda  }{12 \omega} p_7 + \tfrac{k^2}{72(1-2\omega)} p_4 - \tfrac{k^2}{12(1-2\omega)^2} \, p_2 \, {\ln}\left( \tfrac{2}{3} (1+\omega) \right)\right\} R \, \\
& \, + \tfrac{Z^2 \lambda^2 (1 + 20 \omega) }{8 \omega^2}    + \tfrac{Z \lambda \left( (4 + 112 \omega) \Lambda +  k^2 p_5 \right)}{6 \omega}  + \tfrac{k^4 p_3 }{36(1-2\omega)}  
+ \tfrac{k^4\, p_1}{6(1-2\omega)^2}  \, {\ln}\left( \tfrac{2}{3} (1+\omega) \right)
\bigg] \, , 
\end{split}
\ee
where (also for later purposes), we abbreviated 
\be
\begin{array}{lll}
p_1 =  6 - 96 \omega - 48 \omega^2 \, , \qquad & 
p_2 =  65 + 28 \omega + 8 \omega^2 \, , \qquad &
p_3 =  162 - 540 \omega \, , \\
p_4 =  35 - 218 \omega - 352 \omega^2 \, , \qquad &
p_5 =  -2 - 20 \omega \, , \qquad &
p_6 =  1 + 86 \omega + 40 \omega^2 \, , \\
p_7 =  3 + 26 \omega - 40 \omega^2 \, .
\end{array}
\ee
%

\section{Perturbative $\beta$-functions and their fixed points}
%
The $\beta$-functions $\p_t g_i = \beta_{g_i}$ governing the scale-dependence of the coupling constants contained in \eqref{ansatz} can then be read off by comparing the coefficients of the curvature polynomials appearing in \eqref{floweq}. The $\beta$-functions for the marginal couplings are universal in the sense that they do not depend on the regularization scheme. Explicitly, we find
\be\label{beta_universal}
\begin{split}
 \b_\l = &
 - \tfrac{1}{(4\pi)^2}\, \tfrac{133 }{ 10} \, \lambda ^2\, , \qquad 
 \b_\theta = 
 - \tfrac{1}{(4\pi)^2}\, \tfrac{7(56+171 \theta)}{90} \lambda\, , \qquad  
 \b_\omega = 
 - \tfrac{1}{(4\pi)^2}\, \tfrac{(25+1098 \omega+200 \omega ^2)}{60} \lambda \, . 
\end{split}
\ee
Despite our quite different computational approach, this result agrees with earlier computations \cite{deBerredoPeixoto:2004if,Codello:2006in}, 
giving credibility to the off-diagonal heat-kernel methods underlying the URGM. The $\beta$-functions governing the running of Newton's constant
and the cosmological constant are most conveniently written in terms of the dimensionless quantities  $ \tilde{G} = k^2/16\pi Z_k$ and
$\tilde{\Lambda} = k^{-2}\Lambda$:
\be\label{beta_non_universal}
\begin{split}
\beta_{\tilde{G}} = & \, 2 \tilde{G} -  \tfrac{\lambda \Gt}{192 \pi^2 \omega} p_7 - \tfrac{\tilde{G}^2}{12 \pi} 
\left[ \tfrac{1}{6(1-2\omega)} \, p_4 - \tfrac{1}{(1-2\omega)^2} \, p_2 \, {\ln}\left( \tfrac{2}{3} (1+\omega) \right) \right] \, , \\
\beta_{\tilde{\Lambda}} = & \, - 2 \tilde{\Lambda} 
+ \tfrac{\lambda^2 (1+20\omega^2)}{4096 \pi^3 \Gt \omega^2}
+ \tfrac{\Gt}{12 \pi (1-2\omega)^2} (p_1 + \Lt p_2) {\ln}\left( \tfrac{2}{3} (1+\omega) \right) \\ & \,
+ \tfrac{\Gt}{72 \pi (1-2\omega)} (p_3 - \Lt p_4 )
+ \tfrac{\lambda}{192 \pi^2 \omega} (p_5 + \Lt p_6) \, . 
\end{split}
\ee
These are non-universal in the sense that they depend on the chosen regularization scheme. Therefore they are expected to differ from the derivation
\cite{Codello:2006in}, which employs the FRGE with a Type III cutoff. In particular, the log-terms in \eqref{beta_non_universal} are a novel feature in the
Type I cutoff computation. Their appearance can be traced back to the denominators $(P_k - \sigma \Delta)$ appearing in \eqref{prop}, which are absent in the spectrally 
adjusted case.

Owed to their key role for studying the renormalization properties of a theory, we close with a discussion of the FPs $g_i^*$, $(\beta_{g_i}|_{g_i = g_i^*} = 0)$,
of the $\beta$-functions \eqref{beta_universal} and \eqref{beta_non_universal}. The equation $\beta_\lambda(g_i^*) = 0$ has the sole solution $\lambda^* = 0$ and indicates that $\lambda_k$ vanishes logarithmically at high energies. Thus the coupling is asymptotically free.
The remaining equations in \eqref{beta_universal} give rise to the two (well-known) fixed point solutions
\be\label{Fps}
\begin{split}
{\rm FP}_{1,2}:& \qquad \qquad \lambda^*=0\, , \qquad \theta^* = - 171/56 \, , \qquad \omega^*_{1,2} = \{ - 0.00228 \, , \, - 5.47\, \} \, . \\
\end{split}
\ee
Substituting this result into \eqref{beta_non_universal}
we find that only FP$_1$ constitutes a FP of the full system
\be\label{NGFP}
{\rm NGFP}: \qquad \lambda^*=0\, , \quad \theta^* = - 171/56 \, , \quad \omega^* =  - 0.00228  \, , \quad \Lt^* = 0.39 \, , \quad \Gt^* = 2.39 \, .
\ee
This NGFP is UV-attractive in all five couplings. Similarly to the computations \cite{Codello:2006in}, including the quadratic and quartic divergences in the computation of the $\beta$-functions has shifted the couplings $\Lt^*$, $\Gt^*$ to finite values, rendering the theory asymptotically safe instead of asymptotically free. Investigating the fate of FP$_2$, we note that the non-universal $\beta$-functions are well-defined in the region $\omega>-1$ only. This bound can be traced back to the requirement of positivity of the ghost operator \eqref{g2reg}. Since FP$_2$ is not within this bound, it cannot be completed to a FP on the full theory space. We take this as a strong indication that this fixed point is unphysical.

\section{Conclusion and Outlook}

In this paper, we used the exact functional renormalization group equation \cite{Reuter:1996cp} in order to derive the one-loop RG flow of higher derivative gravity based on a novel resummation method inspired by \cite{Barvinsky:1985an}. The main advantage of this technique is that it expresses functional traces of any differential operator, in particular higher-derivative and non-minimal ones, in terms of traces build from Laplace type operators, that are well known and in general easier to compute. Moreover, it allows for a straightforward extension of the one-loop flow to the non-perturbative one. 

Besides the logarithmic singularities seen within dimensional reduction, the functional RG scheme employed above also takes into account quadratic and quartic divergences in the regularization procedure \cite{Codello:2006in,Niedermaier:2009zz}. The latter give rise to a fundamental contribution to the flow of Newtons constant and the cosmological constant, whose UV-behavior is then governed by a non-Gaussian fixed point instead of the Gaussian fixed point seen within dimensional regularization. Furthermore, our regularization scheme strongly suggests that, out of the two known perturbative fixed points \eqref{Fps} for the higher derivative couplings, only one is physically viable. The occurrence of a non-Gaussian fixed point also at the perturbative level provides further hints that gravity may be an asymptotically safe theory.

It would be interesting to study the fate of the NGFP \eqref{NGFP} in a more elaborate computation which also takes the feedback of the running coupling constants appearing on the right-hand-side of the flow equation into account along the lines \cite{Benedetti:2009rx}. We will come back to this point in a future publication \cite{prep}.


\end{document}